# Lemaître's Big Bang

**Jean-Pierre Luminet**

*Aix-Marseille Université, CNRS, Laboratoire d'Astrophysique de Marseille (LAM) UMR 7326*
*& Centre de Physique Théorique de Marseille (CPT) UMR 7332*
*& Observatoire de Paris (LUTH) UMR 8102*
*France*
*E-mail:* `jean-pierre.luminet@lam.fr`

I give an epistemological analysis of the developments of relativistic cosmology from 1917 to 1966, based on the seminal articles by Einstein, de Sitter, Friedmann, Lemaître, Hubble, Gamow and other historical figures of the field. It appears that most of the ingredients of the present-day standard cosmological model, including the acceleration of the expansion due to a repulsive dark energy, the interpretation of the cosmological constant as vacuum energy or the possible non-trivial topology of space, had been anticipated by Georges Lemaître, although his articles remain mostly unquoted.

## 1. Introduction

The full history of Relativistic Cosmology can be divided into 6 periods:

a) An *initial one* (1917-1927), during which the first relativistic universe models were derived in the absence of any observational clue.

b) A *period of development* (1927-1945), during which the cosmological redshifts were discovered and interpreted in the framework of dynamical Friedmann-Lemaître solutions, whose geometrical and mathematical aspects were investigated in more details.

c) A *period of consolidation* (1945-1965), during which primordial nucleosynthesis of light elements and fossil radiation were predicted.

d) A *period of acceptation* (1965-1980), during which the big bang theory triumphed over the rival steady state theory.

e) A *period of enlargement* (1980-1998), when high energy physics and quantum effects were introduced for describing the early universe.

a) The *present period of high precision experimental cosmology*, where the fundamental cosmological parameters are measured with a precision of a few per cent, and new problematics arise such as the nature of the dark energy or cosmic topology.

In this communication I will concentrate on the first three periods to emphasize the prominent contributions of Georges Lemaître, which still subtend the major features of the present-day standard big bang model.

## 2. From static to dynamical universe models

In 1915, Einstein and Hilbert provided the correct field equations for a relativistic theory of gravitation, namely general relativity. Simple cosmological solutions of Einstein's equations can be obtained by assuming homogeneity and isotropy in the matter-energy distribution. This implies that space curvature is on the average constant (*i.e.* it does not vary from point to point, although it may change with time).

The first exact solution was obtained in 1917 by Einstein himself [1], who quite naturally wished to use his brand new theory to describe the structure of the universe as a whole. He assumed that space had a positive curvature, namely the geometry of the hypersphere, and searched for a static solution, *i.e.* in which the average matter density was constant over time, as well as the radius of the hyperspherical space. Einstein expected that general relativity would support this view. However this was not the case. The universe model that he initially calculated did not have a constant radius of curvature: the inexorable force of gravity, acting on each celestial body, had a tendency to make it collapse. The only remedy was to add an *ad hoc* but mathematically coherent term to his original equations. This addition corresponds to some sort of « antigravity », which acts like a repulsive force that only makes itself felt at the cosmic scale. Thanks to this mathematical trick, Einstein's model remained as permanent and invariable as the apparent Universe. The new term, called the *cosmological constant*, has to keep exactly the same value in space and time. Formally, it can take any value, but Einstein fitted it to a specific value $\lambda_E$ in order to constrain the radius $R_E$ of the hypersphere and the matter density $\rho$ to remain constant over time. He thus derived the relation $\lambda_E = 1/R_E^2$.

For Einstein, the fact that space had to be static was a natural assumption since at that time, no astronomical observation indicated that stars had large velocities. In fact, his main motivation was to get a finite space (although without a boundary) and try to fit his solution with Mach's ideas about the origin of inertia.

In the same year, 1917, the Dutch astrophysicist Willem de Sitter [2] derived another model for a static relativistic universe, which was very different from that of Einstein. He assumed that space was positively curved (in fact the projective hypersphere, also called elliptic space, where the antipodal points of the ordinary hypersphere are identified), and empty (in other words, the matter density is zero). As a counterpart, in the absence of matter and therefore of gravity, only a cosmological constant could determine the curvature of space, through the relation $\lambda = 3/R^2$. A strange consequence was that, although the hyperspherical space was assumed to be static (*i.e.* $R$ = constant), the spatial separation between any two test particles had to increase with time. This meant that the cosmological constant has a particular influence on the structure of space: it generates « motion without matter ». However for Einstein, de Sitter's solution reduced to a simple mathematical curiosity, since the real universe indeed has a mass.

In an article of 1922, entitled *On the Curvature of Space* [3], the Russian physicist Alexander Friedmann took the step which Einstein had balked at : he abandoned the theory of a static universe, proposing a dynamical alternative in which space varied with time. As he stated in the introduction, « the goal of this notice is the proof of the possibility of a universe whose spatial curvature is constant with respect to the three spatial coordinates and depend on time, e.g. on the fourth coordinate ». Thus Friedmann assumed a positively curved space (the hypersphere), a time variable matter density $\rho(t)$, and a vanishing cosmological constant. He obtained his famous « closed universe model », with a dynamics of expansion / contraction. Friedmann also derived solutions with non-zero cosmological constant, but pointed out that the term was superfluous. Contrarily to a current opinion, Friedmann's work was not purely mathematical; but he was honest enough to recognize that the available astronomical observations could not support his model : « our information is completely insufficient to carry out numerical calculations and to distinguish which world our universe is. [...] If we set $\lambda = 0$ and $M = 5.10^{21}$ solar masses, the world period becomes of the order 10 billion years ». It was a remarkable prediction, since the most recent estimate for the age of the universe is 13.8 billion years.

Einstein reacted quickly to Friedmann's article. In a short *Note on the work of A. Friedmann 'On the curvature of space'* [4], he argued that « the results concerning the non-stationary world, contained in [Friedmann's] work, appear to me suspicious. In reality it turns out that the solution given in it does not satisfy the field equations ». Of course Friedmann was disappointed. As he could not leave Soviet Union to meet Einstein in Berlin, he wrote an explanatory letter and asked his friend Yuri Krutkov to convince the famous physicist. The mission was apparently successful, since in 1923 Einstein published a still shorter *Note on the work of A. Friedmann 'On the Curvature of Space'* [5], where he recognized an error in his calculations and concluded that « the field equations admit, for the structure of spherically symmetric space, in addition to static solutions, dynamical solutions ». The statement did not mean that Einstein accepted the physical pertinence of dynamical solutions. Indeed, one can find in the original manuscript that the last sentence did not end as in the published version, but with a concluding disavowal « to which it is hardly possible to give a physical meaning » [6].

In 1924, Friedmann studied *On the possibility of a world with constant negative curvature* [7]. He thus assumed a hyperbolic geometry for space, a time varying matter density, and derived the « open » universe model, *i.e.* with a dynamics of perpetual expansion. At the end of his article, he had the first insight on a possibly non-trivial topology of space. Unfortunately, the article remained unnoticed, and Friedmann could never gain the satisfaction to see his theoretical models confronted to cosmological observations: he died prematurely in 1925 after an ascent on a balloon (as he was also a meteorologist).

3. **Lemaître comes into play**

It was precisely the time when the experimental data began to put in question the validity of static cosmological models. For instance, in 1924 the British theorist Arthur Eddington [8] pointed out that, among the 41 spectral shifts of galaxies as measured by Vesto Slipher, 36 were redshifted; he thus favoured the de Sitter cosmological solution while, in 1925, his student, the young Belgian priest Georges Lemaître, proved a linear relation distance-redshift in de Sitter's solution. Contrary to Friedmann, who came to astronomy only three years before his premature death only, Lemaître was closely related to astronomy all his life. He always felt the absolute need for confronting the observational facts and the general relativity theory (adding later considerations from quantum mechanics). He was, for example, much more aware than most of his contemporaries of the experimental status of relativity theory, and that as early as in his years of training. Lemaître was no less a remarkable mathematician, in the domain of fundamental mathematics (see his works on the quaternions or Störmer's problem) as well as in numerical analysis.

The same year 1925, Edwin Hubble proved the extragalactic nature of spiral nebulae [9]. In other words, he confirmed that there existed other galaxies like our own, and the observable universe was larger than previously expected. More important, the radiation from the faraway galaxies was systematically redshifted, which, interpreted as a Doppler effect, suggested that they were moving away from us at great speed. How was it possible?

It was Lemaître who solved the puzzle. In his 1927 seminal paper *Un univers homogène de masse constante et de rayon croissant, rendant compte de la vitesse radiale des nébuleuses extragalactiques*, published (obviously in French) in the Annales de la Société Scientifique de Bruxelles [10], Lemaître calculated the exact solutions of Einstein's equations by assuming a positively curved space (with elliptic, *i.e.* non simply-connected, topology), time varying matter density and pressure, and a non-zero cosmological constant. He obtained a model with perpetual accelerated expansion, in which he adjusted the value of the cosmological constant such as the radius of the hyperspherical space $R(t)$ constantly increased from the radius of the Einstein's static hypersphere $R_E$ at $t = -\infty$. Therefore there was no past singularity and no « age problem ». The great novelty was that Lemaître provided the first interpretation of cosmological redshifts in terms of space expansion, instead of a real motion of galaxies: space was constantly expanding and consequently increased the apparent separations between galaxies. This idea proved to be one of the most significant discoveries of the century.

Using the available astronomical data of the time, Lemaître provided the explicit relation of proportionality between the apparent recession velocity and the distance : « Utilisant les 42 nébuleuses extra-galactiques figurant dans les listes de Hubble et de Strömberg, et tenant compte de la vitesse propre du Soleil, on trouve une distance moyenne de 0,95 millions de parsecs et une vitesse radiale de 600 km/s, soit 625 km/s à $10^6$ parsecs. Nous adopterons donc $R'/R = v/rc = 0,68.10^{-27}$ $cm^{-1}$ (Eq. 24) ». Eq. 24 is exactly what would be called later the Hubble's law.

The fundamental significance of Lemaître's work remained unnoticed. Eddington, his former mentor to whom Lemaître had sent a copy, did not react. When Lemaître met Einstein for the first time at the 1927 Solvay Conference, the famous physicist made favourable technical remarks, but concluded by saying that « from the physical point of view, that appeared completely abominable » [11]. Einstein's response to Lemaître shows the same unwillingness to change his position that characterized his former replies to Friedmann: he accepted the mathematics, but not a physically expanding universe.

In 1929, H. P. Robertson [12] mathematically derived the metrics for all spatially homogeneous universes, but did not realize their physical meaning (his compatriot A. Walker did the same job in 1936, so that in the United States, the Friedmann-Lemaître solutions became unduly called the Robertson-Walker models) Also in 1929, Hubble [13] published the experimental data showing a linear velocity-distance relation $v = Hr$ with $H = 600$ km/s/Mpc. This law was strictly identical to Lemaître's Eq.24, with almost the same proportionality factor, but Hubble did not make the link with expanding universe models. In fact Hubble never read the Lemaître's paper; he interpreted the galaxy redshifts as a pure Doppler effect (due to a proper velocity of galaxies), instead of as an effect of space expansion. However, over the course of the 1920's, spiral galaxies were discovered with redshifts greater than 0,1, which implied recession velocities as large as 30,000 km/s. In 1931, in a letter to de Sitter, Hubble expressed his inability to find a theoretical explanation: « we use the term 'apparent velocities' in order to emphasize the empirical features of the correlation. The interpretation, we feel, should be left to you and the very few others who are competent to discuss the matter with authority. » Also he was not aware that the proportionality factor between redshift and distance, wrongly named the « Hubble constant », was not a constant since it varies with time. Thus it is quite erroneous to claim, as it is often the case, that Hubble is the « father » of the big bang theory. In his popular book of 1936, *The realm of nebulae* [14], the great astronomer honestly recognizes that « the present author is chiefly an observer » and, on the 202 pages of the book, the theoretical interpretation of observations fills only one page (p. 198). Hubble makes reference to Friedmann, Robertson and Milne (who tried to build a Newtonian, non relativistic cosmology), but not to Lemaître.

In 1930, Eddington re-examined Einstein's static model and discovered that, like a pen balanced on its point, it is unstable: with the least perturbation, it begins either expanding or contracting. Thus he called for new searches in order to explain the recession velocities in terms of dynamical space models. Lemaître recalled him that he had already solved the problem in his 1927 article. Eddington, who had not read the paper at the right time, made apologies and promoted the Lemaître's model of expanding space. A new opportunity for the recognition of Lemaître's model arose early in 1930. A discussion between Eddington and De Sitter took place at a meeting of the Royal Astronomical Society in London. They did not know how to interpret the data on the recession velocities of galaxies. Eddington suggested that the problem could be due to the fact that only static models of the universe were hitherto considered, and called for new searches in order to explain the recession velocities in terms of dynamical space models [15].

Having read the report of the meeting of London, Lemaître understood that Eddington and De Sitter posed a problem that he had solved three years earlier. He thus wrote to Eddington to remind him about his communication of 1927 and requested him to transmit a copy to de Sitter. Eddington was somewhat embarrassed. He made apologies and published an important article [16] in which he re-examined the Einstein static model and discovered that, like a pen balanced on its point, it was unstable: any slight disturbance in the equilibrium would start the increase of the radius of the hypersphere; thus he adopted Lemaître's model of the expanding universe - which will be henceforward referred to as the Eddington-Lemaître model.

Eventually, Eddington sponsored the English translation of the 1927 Lemaître's article for publication in the M.N.R.A.S. [17]. An intriguing discrepancy between the original French article and its English translation had already been quoted by various authors (e.g. [18]): the important paragraph discussing the observational data and eq. (24) where Lemaître gave the relation of proportionality between the recession velocity and the distance (in which the determination of the constant that later became known as Hubble's constant appears) was replaced by a single sentence: « From a discussion of available data, we adopt $R'/R = 0,68 \times 10^{-27}$ cm$^{-1}$. » It was found curious that the crucial paragraphs assessing the Hubble law were dropped, so that Lemaître was never recognized as the discoverer of the expansion of the universe. *De facto* Lemaître was eclipsed and multitudes of textbooks proclaim Hubble as the discoverer of the expanding universe, although Hubble himself never believed in such an explanation.

Suddenly, in 2011, a burst of accusations has flared up against Hubble, from the suspicion that a censorship was exerted either on Lemaître by the editor of the M.N.R.A.S. [19] or on the editor by Hubble himself [20] – suspicion based on the « complex personality » of Hubble, who strongly desired to be credited with determining the Hubble constant. The controversy was ended [21] with the help of the Archives Lemaître at Louvain and the Archives of the Royal Astronomical Society: Lemaître himself translated his article, and he chose to delete several paragraphs and notes without any external pressure! On the contrary, he was encouraged to add comments on the subject; but the Belgian scientist, who had indeed new ideas, preferred to publish them in a separate article, published in the same issue of M.N.R.A.S.

### 4. From the Primeval atom to the Hot Big Bang model

Thus, at the beginning of 1931, the expansion of space appeared to be the only coherent explanation to account for the astronomical observations. But the same year when his vision of a dynamic universe was to be accepted by the scientific community, including Eddington, de Sitter and Einstein, Lemaître dared to make an even more outrageous assumption: if the universe is expanding now, must it not have been much smaller and denser at some time in the past? In *The Expanding Universe* [22], he assumed a positively curved space (with elliptic topology), time-varying matter density and pressure, and a cosmological constant such that, starting from a singularity, the Universe first expands, then passes through a phase of « stagnation » during which its radius coasts that of the Einstein's static solution, then starts again in accelerated expansion. This « hesitating model » solved the age problem and provided enough time to form galaxies: « I am led to come around to a solution of the equation by Friedmann where the radius of space starts from zero with an infinite speed, slows and passes by the unstable equilibrium [...] before expanding once again at accelerated speed. It is this period of slowing which seems to me to have played one of the most important roles in the formation of the galaxies and stars. It is obviously essentially connected to the cosmological constant ». Lemaître introduced the revolutionary concept of the « Primeval Atom »: in the distant past the universe must have been so condensed that it was a single entity, which he envisaged as a « quantum of pure energy », referring to the then new discipline of quantum physics. And he poetically described the birth of the Universe: « The atom-world was broken into fragments, each fragment into still smaller pieces [...] The evolution of the world can be compared to a display of fireworks that has just ended: some few red wisps, ashes and smoke. Standing on a cooled cinder, we see the slow fading of the suns, and we try to recall the vanishing brilliance of the origin of the worlds ».

In this Lemaître's *annus mirabilis*, the short note *The beginning of the world from the point of view of quantum theory*, published in Nature [23], can be considered as the chart of the modern big bang theory. Trying to find a link between nebulae and atoms, he applied the latest knowledge about particles and radioactivity: « A comprehensive history of the universe ought to describe atoms in the same way as stars [...] In atomic processes, the notions of space and time are no more than statistical notions: they fade out when applied to individual phenomena involving but a small number of quanta. If the world has begun with a single quantum, the notions of space and time would altogether fail to have any sense at the beginning and would only begin to get some sensible meaning when the original quantum would have been divided in a sufficient number of quanta. If this suggestion is correct, the beginning of the world happened a little before the beginning of space and time. Such a beginning of the world is far enough from the present order of nature to be not at all repugnant. » The radical innovation introduced by Lemaître thus consisted in linking the structure of the universe at large scales with the intimate nature of the atoms, in other words in relating the early universe to quantum mechanics.

Other scientists very poorly received this idea. The fact that Lemaître was a mathematician, allied to his religious convictions (he had been ordained as a priest in 1923), no doubt added to their natural resistance towards the instigation of a new world view. According to Eddington, « the notion of a beginning of the world is repugnant to me », while Einstein considered the primeval atom hypothesis « inspired by the Christian dogma of creation, and totally unjustified from the physical point of view ».

This was an unfair prejudice, because for Lemaître, as he expressed several times, the physical beginning of the world was quite different from the metaphysical notion of creation. And for the priest-physicist, science and religion corresponded to separate levels of understanding. It is interesting to point out that the manuscript (typed) version of Lemaître's article, preserved in the Archives Lemaître at the Université of Louvain, ended with a sentence crossed out by Lemaître himself and which, therefore, was never published. Lemaître initially intended to conclude his letter to *Nature* by « I think that every one who believes in a supreme being supporting every being and every acting, believes also that God is essentially hidden and may be glad to see how present physics provides a veil hiding the creation ». This well reflected his deep theological view of a hidden God, not to be found as the Creator in the beginning of the universe. But before sending his paper to the journal, Lemaître probably realized that such a reference to God could mislead the readers and make them think that his hypothesis gave support to the Christian notion of God. Unfortunately, it is precisely what they did.

Einstein had also a bad opinion of the cosmological constant, that he considered as the « greatest blunder of his life ». It is probably the reason why, in the new relativistic model that he proposed in 1932 with de Sitter [24] - a Euclidean model with uniform density that expanded eternally - the term disappeared. The authors did not even make reference to Friedmann and Lemaître's works, and after that, Einstein forgave research in cosmology...

Unfortunately, due to Einstein's authority, this over-simplified solution became the standard model of cosmology for the next 60 years. However Lemaître kept his original views. In 1933 he published another fundamental article about cosmology, galaxy formation, gravitational collapse and singularities [25]. In that paper of 1933, Lemaître found a new solution of Einstein's equations, known as the « Lemaître-Tolman » model, which is more and more frequently used today for considering structure formation and evolution in the real Universe within the exact (*i.e.* non-perturbative) Einstein theory. In the less known *Evolution of the expanding universe* published in 1934 [26], he had a first intuition of a cosmic background temperature at a few Kelvins: « If all the atoms of the stars were equally distributed through space there would be about one atom per cubic yard, or the total energy would be that of an equilibrium radiation at the temperature of liquid hydrogen. » He also interpreted for the first time the cosmological constant as vacuum energy: « The theory of relativity suggests that, when we identify gravitational mass and energy, we have to introduce a constant. Everything happens as though the energy in vacuo would be different from zero. In order that motion relative to vacuum may not be detected, we must associate a pressure $p = -\rho c^2$ to the density of energy $\rho c^2$ of vacuum. This is essentially the meaning of the cosmological constant $\lambda$ which corresponds to a negative density of vacuum $\rho_0$ according to $\rho_0 = \lambda c^2/4\pi G$

~ $10^{-27}$ *gr.cm$^{-3}$* ». Such a result will be rediscovered only in 1967 by Sakharov [27] on the basis of quantum field theory, and is now considered as one of the major solutions of the so-called « dark energy problem ».

By 1950, when Lemaître published a summary, in English, of his theory, entitled *The Primeval Atom: An Essay on Cosmogony* [28], it was thoroughly unfashionable. Two years previously the rival theory of a « steady state » universe, supported principally by Thomas Gold, Hermann Bondi and Fred Hoyle[29], had met with widespread acclaim. Their argument was that the universe had always been and would always be as it is now, that is was eternal and unchanging. In order to obtain what they wanted, they assumed an infinite Euclidean space, filled with a matter density constant in space and time, and a new « creation field » with negative energy, allowing for particles to appear spontaneously from the void in order to compensate the dilution due to expansion. Seldom charitable towards his scientific adversaries, Fred Hoyle made fun of Lemaître by calling him « the big bang man ». In fact he used for the first time the expression « big bang » in 1948, during a radio interview.

The term, isolated from its pejorative context, became part of scientific parlance thanks to a Russian-born American physicist George Gamow, a former student of Alexander Friedmann. Hoyle therefore unwittingly played a major part in popularizing a theory he did not believe in; he even brought grist to the mill of big bang theory by helping to resolve the question why the universe contained so many chemical elements. Claiming that all the chemical elements were formed in stellar furnaces, he was contradicted by Gamow and his collaborators[30]. The latter took advantage of the fact that the early universe should have been very hot. Assuming a primitive mixture of nuclear particles called *Ylem*, a Hebrew term referring to a primitive substance from which the elements are supposed to have been formed, they were able to explain the genesis of the lightest nuclei (deuterium, helium, and lithium) during the first three minutes of the Universe, at an epoch when the cosmic temperature reached 10 billion degrees. Next they predicted that, at a later epoch, when the Universe had cooled to a few thousand degrees, it suddenly became transparent and allowed light to escape for the first time. Alpher and Hermann [31] calculated that one should today receive an echo of the big bang in the form of blackbody radiation at a fossil temperature of about 5 K. Their prediction did not cause any excitement. They refined their calculations several times until 1956, without causing any more interest; no specific attempt at detection was undertaken.

In the middle of the 1960's, at Princeton University, the theorists Robert Dicke and James Peebles studied oscillatory universe models in which a closed universe in expansion-contraction, instead of being infinitely crushed in a big crunch, passes through a minimum radius before bouncing into a new cycle. They calculated that such a hot bounce would cause blackbody radiation detectable today at a temperature of 10 K. It was then that they learned that radiation of this type had just been detected, at the Bell Company laboratories in New Jersey. There, the engineers Arno Penzias and Robert Wilson had been putting the finishing touches on a radiometer dedicated to astronomy, and they had found a background noise that was higher than expected. After subtracting the antenna noise and absorption by the atmosphere, there remained an excess of 3.5 K. This background noise had to be of cosmic origin: it was the fossil radiation. The teams of the Bell Company and Princeton University published their articles separately in the same issue of July 1965 of the *Astrophysical Journal* [32]. Penzias and Wilson only gave the results of their measurements, while Dicke, Peebles, Roll and Wilkinson gave their cosmological interpretation.

None of them mentioned the predictions of Alpher and Hermann, still less those of Lemaître. The latter died in 1966, a few weeks after his assistant informed him about the discovery of the fossil radiation (Lemaître is supposed to have commented « I am glad now, we have the proof »). Gamow also died in 1968 without being recognized for his predictions. Alpher and Herman were almost forgotten. Penzias and Wilson gained the Nobel Prize in physics in 1978. Nevertheless, at the moment of their discovery, they believed instead in the theory of continuous creation, rival to that of the big bang, while their detection of the fossil radiation practically signalled the death sentence of the steady state model.

After half a century of rejection, Lemaître's primeval atom, in the guise of the catchphrase « big bang theory », had at last been accepted by theoretical physicists.